# Optimal Distribution System Restoration via Tractable Modeling of Decision-Dependent Interruption Cost and Cold Load Pickup

Wei Wang, Minwu Chen, Hongbin Wang, Gaoqiang Peng, Hongzhou Chen

*Abstract*—Developing optimized restoration strategies for power distribution systems (PDSs) is critical to enhancing resilience. Prior knowledge of customer interruption cost (CIC) and load restoration behaviors, particularly cold load pickup (CLPU), is essential for effective decision-making. However, both CIC and CLPU are reciprocally influenced by the realized customer interruption duration (CID), making them decision-dependent and challenging to model, especially given the limited understanding of their underlying physical mechanisms. This paper proposes a novel and tractable modeling approach to capture the varying patterns of CIC and CLPU with CID—patterns derived from data that reflect observable surface-level correlations rather than underlying mechanisms—thereby enabling practical surrogate modeling of these decision-dependent factors. Specifically, quadratic functions are employed to model the increasing rate of CIC with respect to CID according to data fitting results. For CLPU, several defining characteristics are extracted and modeled in a piecewise linear form relative to CID, and the actual restored load accounting for CLPU is subsequently reconstructed. Building on these models, a PDS restoration optimization framework is developed, incorporating mobile energy storage systems (MESSs) and network reconfiguration strategies. Case studies validate the effectiveness of the proposed approach and highlight MESS's unique potential in accelerating CLPU-related restoration.

## Nomenclature

**Sets**

| | |
|---|---|
| $\mathcal{T}$ | Set of time spans. |
| $\mathcal{N}_X$ | Set of initially interrupted loads. |
| $\mathcal{K}_\alpha, \mathcal{K}_D, \mathcal{K}_t$ | Sets of binary bit positions for decomposing $\alpha_{i,t}$, $D_{\text{dc},i}$, and $t_{2,i}$, respectively. |
| $\mathcal{M}$ | Set of MESSs. |
| $\mathcal{N}_M$ | Set of nodes accessible to MESSs. |
| $\mathcal{N}/\mathcal{L}$ | Set of power nodes/branches of DS. |
| $\mathcal{L}_{\text{out},t}$ | Set of damaged branches during time span $t$. |

**Variables**

| | |
|---|---|
| $\delta_{i,t}$ | Binary; 1 if load $i$ is restored during time span $t$, 0 otherwise. |
| $\alpha_{i,t}$ | Integer; CID of load $i$ until the end of time span $t$. |
| $A_{i,t}^{\text{fun}}$ | Estimated increasing rate of CIC for load $i$ during time span $t$. |
| $\lambda_{k,i}^{(\alpha)}, \lambda_{k,i}^{(D)}, \lambda_{k,i}^{(t)}$ | Binary; auxiliary variables for decomposing $\alpha_{i,t}$, $D_{\text{dc},i}$, and $t_{2,i}$, respectively. |
| $z_{k,i,t}^{(-)}, z_{k,i}^{(-)}, \beta_{i,t}, W_{\sim,i,t}$ | Auxiliary variables for linearizing the bilinear terms. |
| $F_i$ | Peak magnitude of CLPU for load $i$ upon restoration. |
| $D_{\text{pk},i}, D_{\text{dc},i}$ | Integer; peak and decay durations of CLPU for load $i$ upon restoration, respectively. |
| $d_{\text{intr},i}$ | Accumulated CID before restoration of load $i$. |
| $\kappa_{F,i}/\kappa_{\text{pk},i}/\kappa_{\text{dc},i}$ | Binary; 1 if $F_i/D_{\text{pk},i}/D_{\text{dc},i}$ is within the segment prior to saturation, respectively, 0 otherwise. |
| $u_{1,i,t}, u_{2,i,t}, u_{3,i,t}$ | Binary; indicators for tracking the CLPU process for load $i$. |
| $\Delta f_i$ | Decay rate during the CLPU for load $i$. |
| $\xi_i$ | Binary; 1 if $D_{\text{dc},i} \neq 0$, 0 otherwise. |
| $t_{1,i}, t_{2,i}, t_{3,i}$ | Timestamps for flagging the CLPU process. |
| $P_{L,i,t}, Q_{L,i,t}$ | Actual active and reactive power loads of load $i$ during time span $t$ incorporating CLPU. |
| $x_{j,i,t}$ | Binary variable; 1 if MESS $j$ is parked at node $i$ during time span $t$, 0 otherwise. |
| $v_{j,i,t}$ | Binary variable; 1 if MESS $j$ is traveling to node $i$ during time span $t$, 0 otherwise. |
| $ch_{j,i,t}/dch_{j,i,t}$ | Binary; 1 if MESS $j$ charges/discharges at node $i$ during time span $t$, 0 otherwise. |
| $P_{c,j,i,t}/P_{d,j,i,t}$ | Active charging/discharging power of MESS $j$ at node $i$ during time span $t$. |
| $soc_{j,t}$ | State of charge of MESS $j$ at the end of time span $t$. |
| $f_{ii',t}^{i_x}$ | Directed flow of commodity $i_x$ from nodes $i$ to $i'$ during time span $t$. |
| $\lambda_{ii',t}$ | Binary; 1 if arc $(i, i')$ is included in the directed fictitious spanning tree during time span $t$, 0 otherwise. |
| $\mu_{ii',t}$ | Binary; 1 if branch $(i, i')$ is included in the fictitious spanning tree during time span $t$, 0 otherwise. |
| $\upsilon_{ii',t}$ | Binary variable; 1 if branch $(i, i')$ of the DS is closed during time span $t$, 0 otherwise. |
| $P_{\text{DG},i,t}/Q_{\text{DG},i,t}$ | Active/reactive power output of distributed generating unit at node $i$ during time span $t$. |
| $P_{\text{IN},i,t}/Q_{\text{IN},i,t}$ | Active/reactive power injected by MESSs into node $i$ during time span $t$. |
| $P_{i'i,t}/Q_{i'i,t}$ | Active/reactive power flow on branch ($i'$, $i$) from node $i'$ to node $i$ during time span $t$. |
| $V_{i,t}^2$ | Squared voltage magnitude at node $i$ during time span $t$. |

*Parameter*

| | |
|---|---|
| $\Delta t$ | Length of unit time span. |
| $a_{i*}$, $b_{i*}$, $c_{i*}$ | Quadratic coefficients for increase rate of CIC versus interruption duration for load type $i*$, to which load $i$ belongs. |
| $k_{F,i*}$, $c_{F,i*}$, $k_{pk,i*}$, $c_{pk,i*}$, $k_{dc,i*}$, $c_{dc,i*}$ | Linear coefficients for peak magnitude, peak duration, and decay duration of CLPU versus interruption duration for load type $i*$. |
| $\alpha_{\max}$ | An estimated maximum interruption duration. |
| $a_{i,0}$ | Interruption duration for load $i$ preceding the scheduling. |
| $(\hat{d}_{F,i*},\hat{F}_{i*})/$ $(\hat{d}_{pk,i*},\hat{D}_{pk,i*})/$ $(\hat{d}_{dc,i*},\hat{D}_{dc,i*})$ | Saturation point of the relationship between peak magnitude/peak duration/ decay duration and interruption duration for load type $i*$. |
| $\varepsilon_{pk}/\varepsilon_{dc}$ | Threshold for rounding $D_{pk,i}/D_{dc,i}$. |
| $p_{L,i,t}$ | Original active power load of load $i$ during time span $t$. |
| $P_{c,\max,j}/$ $P_{d,\max,j}$ | Maximum active charging/discharging power of MESS $j$. |
| $Q_{\max,j}$ | Maximum reactive power output of MESS $j$. |
| $e_{c,j}/e_{d,j}$ | Charging/discharging efficiency for MESS $j$. |
| $E_j$ | Energy capacity of MESS $j$. |
| $soc_{j,\min}/$ $soc_{j,\max}$ | Minimum/maximum value of state of charge for MESS $j$. |
| $P_{DG,\max,i}$ $/Q_{DG,\max,i}$ | Maximum active/reactive power output of distributed generating unit at node $i$. |
| $\eta_i$ | The coefficient about power factor of load $i$. |
| $r_{i'i}/x_{i'i}$ | Resistance/reactance of branch ($i'$, $i$). |
| $V_{i,\min}/V_{i,\max}$ | Lower/upper bound of voltage magnitude at node $i$. |
| $S_{i'i}$, $S_j$, $S_{DG,i}$ | Apparent power capacity of branch ($i'$, $i$), MESS $j$, or the unit at node $i$. |
| $\sigma_1, \sigma_2, \sigma_3$ | Penalty coefficients for resources scheduling. |

## I. Introduction

ENHANCING the resilience of power distribution systems (PDSs) has become an urgent priority in both practice and research, driven by the rising frequency of extreme climate events and the global shift of many end-use energy demands toward electrification due to decarbonization efforts [1], [2]. A resilient PDS should, despite the various definitions of resilience, be well-prepared for disasters and capable of quickly restoring from them.

With specific regard to restoration of PDS, recent decades have witnessed a surge in studies that build optimization models to explore the utilization of distributed energy resources (DERs) alongside network reconfiguration, forming microgrids with dynamic boundaries to quickly restore interrupted loads [3]-[5]. The flexibility of this process is particularly enhanced by incorporating the routing of mobile DERs, which include not only dedicated mobile energy storage systems (MESSs) and generators [6]-[8], but also electric buses and ships [9], [10]. However, despite existing innovative efforts, challenges persist in optimization modeling, particularly concerning decision-dependent factors, based on the authors' knowledge and literature review.

Customer interruption cost (CIC) is undoubtedly a key component of the objective function, as it is crucial for prioritizing the sequence of load restoration and guiding the overall process. While significant attention has been focused on constraint modeling for the operational behaviors of emergency resources and PDS, far less consideration has been given to accurately representing CIC. Most studies incorporate a fixed CIC per unit time (*e.g.*, in $/kW·h) into their objective functions, assuming that cumulative CIC (*e.g.*, in $/kW) increases linearly with customer interruption duration (CID) [11], [12]. However, customer surveys have revealed a nonlinear relationship between CIC – whether per unit time or cumulative – and CID [13]-[15], the latter depending on the restoration decisions that need to be optimized. Obviously, accurate CIC is essential for making objective decisions about which loads to restore first during each time step, but this is challenging due to the lack of an analytical model linking CIC to CID; even if developed, such a model would likely be too complex to incorporate into optimization. Consequently, we are compelled to turn to representing CIC based on the limited statistical data we have, such as that from CIC surveys – an important pragmatic task for power utilities [16].

Beyond CIC, another critical decision-dependent factor is the restoration behavior of thermostatically-controlled loads (TCLs), known as the cold load pickup (CLPU) phenomenon, which is becoming increasingly significant with the growing penetration of TCLs in power systems driven by social development [17]. In many Chinese cities, air conditioning in summer has accounted for up to 60% of the peak load [18]. CLPU primarily stems from load inrush and the simultaneous start-up of numerous appliances due to loss of load diversity [19], with the latter being more prolonged and critical in restoration optimization [20]. It results in load surges beyond normal levels, characterized by factors such as peak magnitude, duration, and decay time, depending on elements including CID, ambient temperature, and load composition [19], [21]. While the latter can be accessed via situational awareness, CID remains undetermined until the optimal restoration decisions are made, making CLPU inherently decision-dependent. To ensure the effectiveness and feasibility of the final optimized decisions, we must first understand the extent of CLPU that will occur when restoring the interrupted loads at certain times. Therefore, it is essential to incorporate the relationship between CLPU and CID – considering the latter as a variable to be optimized – into the restoration optimization modeling process. However, similar to CIC, this too presents challenges due to the lack of a computationally tractable model for the CLPU-CID relationship.

Despite notable efforts in a few studies to model the TCL population behaviors from the probabilistic perspective, *e.g.*, using Fokker-Planck partial differential equations [22], these methods still rely on numerical solutions, making them hard to embed in optimization models. As a result, we are currently resorting to a data-driven approach, for example, exhaustively accessing CLPU-related load data across all CID values via simulations or historical records, upon which we base the construction of the restoration optimization model, as seen in recent innovative studies [23]-[25]. However, the current methods still face inherent limitations. First, simulation-based approaches require simulating the physical behaviors of TCL population across the full range of undetermined CID values over the restoration horizon. It is a daunting task, particularly when conducting precise simulations based on computational fluid dynamics (CFD), finite-element analysis, or specialized softwares like EnergyPlus$^{TM}$ [26]-[28], which, though high-fidelity, are computation-intensive. The enormity of this task is evident when attempting exhaustive pre-simulations across all CID values and environmental conditions like temperature, or even when deferred until restoration decisions are imminent – though other factors are clarified, simulating across all CID values remains unavoidable, thus delaying critical decisions. In addition, for record-based approaches, ensuring records cover all CID values is prohibitively challenging due to monitoring limitations, infrequent blackouts, and difficulties with real-world outage experiments.

In light of the challenges in modeling either CIC or CLPU, the concept of data-driven optimization offers an insightful approach. The key idea in this emerging field is constructing surrogate models, also known as metamodels, to approximate objective or constraint functions that would otherwise have complex forms or require computation-intensive simulations or experiments [29], [30]. One major advantage is that these surrogate models can be trained on a limited amount of data, rather than requiring the entire dataset. This means we can be freed from the need for exhaustive simulations or records for each CID value, allowing restoration optimization to be carried out even with limited available CIC or CLPU data. In addition, the advantages also include the ability of well-trained surrogate models (*e.g.*, through data fitting) to reduce the impact of data quality issues, such as noise or outliers [31], [32], thereby mitigating the overfitting risk that may arise when optimization is directly based on original data. Therefore, leveraging these insights, this paper primarily focuses on building tractable surrogate models for decision-dependent CIC and CLPU, which underpin the formulation of optimal PDS restoration problem. These models, though expressed in analytical form, do not represent the actual physical relationships between variables, but rather capture patterns present in the data. To the best of our knowledge, no similar research has been conducted to date. The specific contributions are threefold:

*1)* We characterize the decision-dependency of CIC by defining it as the variable rate at which CIC increases with CID. Based on the results of CIC-CID data fitting, quadratic functions are formulated to represent this dependency and subsequently linearized precisely into tractable constraints.

*2)* The decision-dependent nature of CLPU is captured by extracting its varying characteristics with CID, including peak load magnitude, peak and decay durations, and decay rate. These variations are modeled as piecewise linear functions, and the actual load accounting for CLPU is subsequently reconstructed based on these characteristics.

*3)* Building on the prior models, an optimization framework for PDS restoration is developed, integrating scheduling of MESSs and network reconfiguration. The unique potential of MESSs to mitigate CLPU challenges is demonstrated through their capability of providing "dynamic rating" services across isolated islands, enabling short-term accommodation of CLPU load surges and facilitating enhanced load restoration.

The rest of this paper is organized as follows. Section II and Section III derive the model formulations for CIC and CLPU, respectively; Section IV presents the comprehensive optimization model for PDS restoration; Section V conducts case studies; and Section VI concludes the paper.

## II. MODELING OF DECISION-DEPENDENT CIC

### A. Relationship Between CIC and CID

The electric power industry has spent decades researching CIC through surveys, with contributions from countries like those in North America and Europe [14], [15], [33]. Among the representative outcomes is the ICE Calculator (*https://icecalculator.com/home*), a tool for CIC estimation, developed based on extensive survey datasets and refined through multiple iterations, with support from institutions like the U.S. Department of Energy and Lawrence Berkeley National Laboratory. Using the ICE Calculator, we conducted comprehensive testing to obtain sufficient CIC estimations across a wide range of CID conditions and fitting analysis uncovered a strong quadratic relationship between CID and the increasing rate of CIC, a finding that, to the best of our knowledge, has not been previously reported. Some of the generated samples and their fitting results are depicted in Fig. 1, where California was chosen as the region for estimation, with other parameters set to ICE Calculator defaults, such as household income and industry percentage.

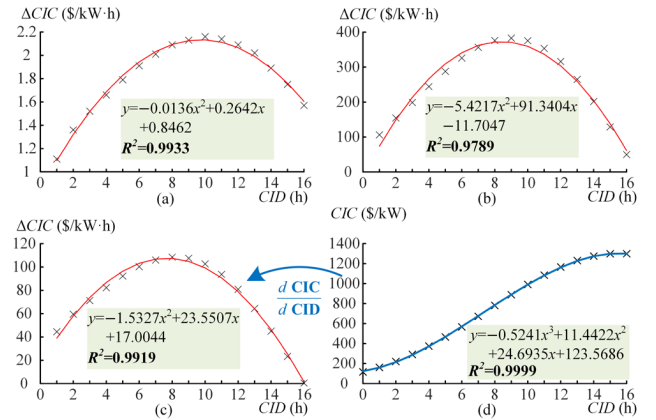

**Fig. 1.** Estimated increasing rate of CIC under varying CID and polynomial fitting results for (a) residential, (b) small C&I, and (c) large C&I customers, respectively; and (d) cumulative CIC and fitting result for large C&I customers.

The quadratic relationship between the increasing rate of CIC and CID, revealed by least squares fitting results such as those in Fig. 1(a)-(c), suggests that, for example, after an 8-

hour outage, the marginal cost of an additional hour is typically higher than after 16 hours without power. This effect is even more pronounced for commercial and industrial (C&I) customers. Additionally, we also performed polynomial fitting on the generated cumulative CIC data, showing a strong fit with a cubic function, as illustrated in Fig. 1(d). This further supports our findings, as the derivative of the cumulative CIC function with CID directly corresponds to the increasing rate. Although higher-order polynomials may offer a better fit, they often lead to overfitting and are generally discouraged in surrogate modeling for data-driven optimization [29]. In addition, they complicate conversion into tractable constraints, such as linear ones, as will be addressed subsequently.

*B. Formulation of Linear Constraints*

Regardless of the intricate actual relationship between CIC and CID, the data has suggested it can be approximated by a quadratic function, upon which we can build its surrogate model for the final optimization. First, we define $\alpha_{i,t}$ as a counter to track the CID that has already been experienced:

$$-M \cdot \delta_{i,t} \leq \alpha_{i,t} - (\alpha_{i,t-1} + 1) \leq 0 \,,\, \forall i \in \mathcal{N}_X, t \in \mathcal{T} \quad (1a)$$

$$0 \leq \alpha_{i,t} \leq \alpha_{\max} \cdot (1 - \delta_{i,t}) \,,\, \forall i \in \mathcal{N}_X, t \in \mathcal{T} \quad (1b)$$

Additionally, $\alpha_{i,0}$ is set to the initial CID for load $i$ at the beginning of the current restoration horizon.

Based on the previous subsection, the increasing rate of CIC, which varies quadratically with the ongoing interruption and becomes 0 once power is restored, can be expressed by

$$0 \leq A_{i,t}^{\text{fun}} \leq M \cdot (1 - \delta_{i,t}) \,,\, \forall i \in \mathcal{N}_X, t \in \mathcal{T} \quad (2a)$$

$$-M \cdot \delta_{i,t} \leq A_{i,t}^{\text{fun}} - \left[ a_{i*} \cdot (\alpha_{i,t} \cdot \Delta t)^2 + b_{i*} \cdot (\alpha_{i,t} \cdot \Delta t) + c_{i*} \right] \quad (2b)$$
$$, \forall i \in \mathcal{N}_X, t \in \mathcal{T}$$

$$\varepsilon - M \cdot \delta_{i,t} \leq A_{i,t}^{\text{fun}} \,,\, \forall i \in \mathcal{N}_X, t \in \mathcal{T} \quad (2c)$$

Particularly, as CID increases, the quadratic function may intersect the x-axis, as shown in Fig. 1(c) when CID exceeds 16 h, leading to an unrealistic negative increasing rate of CIC. In such cases, we set the rate to 0, indicating that CIC no longer increases. Instead of introducing cumbersome binary variables to explicitly represent this piecewise function, we adopt a simpler approach facilitated by the optimization mechanism: we formulate (2b) and (2c) to enforce $A_{i,t}^{\text{fun}}$ to equal the greater value between the current CID-related functional CIC and $\varepsilon$ when load $i$ is interrupted, given that the objective (21) contains "min $A_{i,t}^{\text{fun}}$". $\varepsilon$ is a small positive value close to 0 but cannot be 0, as this could result in the load not being restored once its CID exceeds the intersecting point.

Due to the negative $a_{i*}$, as shown in Fig. 1, constraint (2b) is non-tractable due to non-convexity after LP relaxation, preventing its conversion to a tractable form like a second-order cone. Therefore, we replace $(\alpha_{i,t})^2$ with $\beta_{i,t}$ and, given the integrality of $\alpha_{i,t}$, impose the following constraints to guarantee the exactness of replacement [34]. $\lambda_{k,i,t}^{(\alpha)}$ represents the value of the $k$-th bit in the binary representation of $\alpha_{i,t}$.

$$\alpha_{i,t} = \sum_{k=0}^{\lfloor \log_2 \alpha_{\max} \rfloor} 2^k \cdot \lambda_{k,i,t}^{(\alpha)} \,,\, \forall i \in \mathcal{N}_X, t \in \mathcal{T} \quad (3a)$$

$$-\alpha_{\max} \cdot \left(1 - \lambda_{k,i,t}^{(\alpha)}\right) \leq z_{k,i,t}^{(\lambda \cdot \alpha)} - \alpha_{i,t} \leq 0 \,,\, \forall k \in \mathcal{K}_\alpha, i \in \mathcal{N}_X, t \in \mathcal{T} \quad (3b)$$

$$0 \leq z_{k,i,t}^{(\lambda \cdot \alpha)} \leq \alpha_{\max} \cdot \lambda_{k,i,t}^{(\alpha)} \,,\, \forall k \in \mathcal{K}_\alpha, i \in \mathcal{N}_X, t \in \mathcal{T} \quad (3c)$$

$$\beta_{i,t} = \sum_{k=0}^{\lfloor \log_2 \alpha_{\max} \rfloor} 2^k \cdot z_{k,i,t}^{(\lambda \cdot \alpha)} \,,\, \forall i \in \mathcal{N}_X, t \in \mathcal{T} \quad (3d)$$

Then, (2b) can be reformulated as the linear constraint (4a).

$$-M \cdot \delta_{i,t} \leq A_{i,t}^{\text{fun}} - \left[ a_{i*} \cdot (\Delta t)^2 \cdot \beta_{i,t} + b_{i*} \cdot \Delta t \cdot \alpha_{i,t} + c_{i*} \right] \quad (4)$$

Additionally, the following constraints can be added as cuts to tighten the feasible region of the LP relaxation by forming McCormick Envelops for the quadratic term $(\alpha_{i,t})^2$, some parts of which are already implicit in (3).

$$\beta_{i,t} \geq 2 \cdot \alpha_{\max} \cdot \alpha_{i,t} - \alpha_{\max}^2 \,,\, \forall i \in \mathcal{N}_X, t \in \mathcal{T} \quad (5)$$

### III. MODELING OF DECISION-DEPENDENT CLPU

*A. Relationships Between CLPU Characteristics and CID*

While the dependency of CLPU characteristics (*e.g.*, peak magnitude and duration) on CID has gained attention, research on quantitatively describing these patterns remains limited. Nevertheless, a few studies have attempted to approximate this dependency. For example, linear relationships between CLPU peak magnitude and peak duration with CID were derived as early as in [35], albeit based on many assumptions, and have since been corroborated by recent studies [36], [37]. Since this paper primarily focuses on mathematical and especially tractable modeling rather than directly examining the variation patterns of CLPU characteristics, we build on the foundation of these existing studies to support our work.

Specifically, we adopt a piecewise linear approximation for the CLPU curve, as in [38] and shown in Fig. 2(a). Compared to the traditional exponential decay model [20], [21], this approach tends to overestimate the extent of CLPU, providing an additional safety margin for the restoration process while also simplifying modeling and computational efforts. Three key characteristics are then selected to represent the CLPU process: peak magnitude, peak duration, and decay duration. Finally, we follow the evidence and data from [35]-[37] and assume a two-stage linear relationship with saturation between these characteristics and CID, as shown in Fig. 2(b). Based on this, we will derive constraints to form our model for CLPU. As new findings on the variation patterns of CLPU with CID emerge, the modeling process can be updated accordingly, *e.g.*, by adopting a piecewise linear function with more segments in Fig. 2(b) to capture more complex patterns.

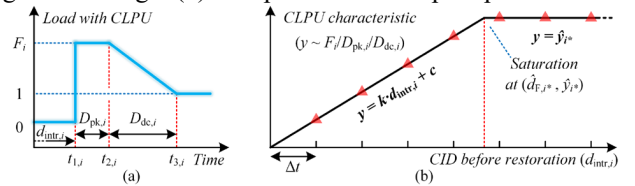

**Fig. 2.** (a) Load modeling with CLPU, where the magnitude is taken as a multiple of the original load; (b) approximated relationships between CLPU characteristics and CID.

## B. Formulation of Linear Constraints

### 1) Key characteristics

First, the variable $d_{\text{intr},i}$ is introduced to represent the final CID before restoration, as formulated by

$$d_{\text{intr},i} = \sum_{t \in \mathcal{T}} (1 - \delta_{i,t}) + \alpha_{i,0}, \quad \forall i \in \mathcal{N}_X \quad (6)$$

Based on Fig. 2(b), we derive (7a)–(7c) to represent the relationship between peak magnitude $F_i$ and CID, as depicted by the points marked with triangles.

$$-M \cdot \kappa_{F,i} \leq d_{\text{intr},i} - \lceil \hat{d}_{F,i} \rceil \leq M \cdot (1 - \kappa_{F,i}) - 1, \quad \forall i \in \mathcal{N}_X \quad (7a)$$

$$-M \cdot (1 - \kappa_{F,i}) \leq F_i - (k_{F,i^*} \cdot d_{\text{intr},i} + c_{F,i^*}) \leq M \cdot (1 - \kappa_{F,i}), \quad \forall i \in \mathcal{N}_X \quad (7b)$$

$$-\hat{F}_{i^*} \cdot \kappa_{F,i} \leq F_i - \hat{F}_{i^*} \leq 0, \quad \forall i \in \mathcal{N}_X \quad (7c)$$

The relationships for peak duration $D_{\text{pk},i}$ and decay duration $D_{\text{dc},i}$ can be similarly expressed, which are converted into integers (i.e., multiples of time step $\Delta t$) to facilitate their integration into the discrete-time restoration optimization scheme. The parameter $\varepsilon \in (0,1)$ is introduced to adjust the rounding threshold – rounding down if the fractional part is below $\varepsilon$, and rounding up otherwise. Therefore, we derive (8a)–(8d) to represent the relationship for $D_{\text{pk},i}$, with similar expressions applicable to $D_{\text{dc},i}$.

$$-M \cdot \kappa_{\text{pk},i} \leq d_{\text{intr},i} - \lceil \hat{d}_{\text{pk},i} \rceil \leq M \cdot (1 - \kappa_{\text{pk},i}) - 1, \quad \forall i \in \mathcal{N}_X \quad (8a)$$

$$D_{\text{pk},i} - (k_{\text{pk},i^*} \cdot d_{\text{intr},i} + c_{\text{pk},i^*} + 1 - \varepsilon_{\text{pk}}) \leq M \cdot (1 - \kappa_{\text{pk},i}), \quad \forall i \in \mathcal{N}_X \quad (8b)$$

$$-M \cdot (1 - \kappa_{\text{pk},i}) \leq D_{\text{pk},i} - (k_{\text{pk},i^*} \cdot d_{\text{intr},i} + c_{\text{pk},i^*} - \varepsilon_{\text{pk}}), \quad \forall i \in \mathcal{N}_X \quad (8c)$$

$$-M \cdot \kappa_{\text{pk},i} \leq D_{\text{pk},i} - \lceil \hat{D}_{\text{pk},i^*} - \varepsilon_{\text{pk}} \rceil \leq 0, \quad \forall i \in \mathcal{N}_X \quad (8d)$$

### 2) Decay rate

As a necessary step in load modeling, the CLPU decay rate is derived as follows, based on the assumption in Fig. 2(a):

$$\text{if } D_{\text{dc},i} \neq 0 \Rightarrow \Delta f_i \cdot D_{\text{dc},i} = F_i - 1$$

Given that $D_{\text{dc},i}$ is an integer, we can perform the similar binary decomposition as in (3a):

$$D_{\text{dc},i} = \sum_{k=0}^{\lfloor \log_2 \lceil \hat{D}_{\text{dc},i} \rceil \rfloor} 2^k \cdot \lambda_{k,i}^{(D)}, \quad \forall i \in \mathcal{N}_X \quad (9)$$

Next, a binary variable $\xi_i$ is introduced to indicate the condition $D_{\text{dc},i} \neq 0$, which can be expressed as $\xi_i = \lambda_{0,i}^{(D)} \vee \lambda_{1,i}^{(D)} \vee \lambda_{2,i}^{(D)} \vee \ldots$ and further formulated by the following constraints:

$$\xi_i \geq \lambda_{k,i}^{(D)}, \quad \forall k \in \mathcal{K}_D, i \in \mathcal{N}_X \quad (10a)$$

$$\xi_i \leq \sum_{k=0}^{\lfloor \log_2 \lceil \hat{D}_{\text{dc},i} \rceil \rfloor} \lambda_{k,i}^{(D)}, \quad \forall i \in \mathcal{N}_X \quad (10b)$$

Then, similarly to (3), the consequent (or right-hand side) of the logical condition can be expressed by the following linear constraints, which impose the correct decay rate.

$$-M \cdot (1 - \lambda_{k,i}^{(D)}) \leq z_{k,i}^{(\lambda \cdot \Delta f)} - \Delta f_i \leq 0, \quad \forall k \in \mathcal{K}_D, i \in \mathcal{N}_X \quad (11a)$$

$$0 \leq z_{k,i}^{(\lambda \cdot \Delta f)} \leq M \cdot \lambda_{k,i}^{(D)}, \quad \forall k \in \mathcal{K}_D, i \in \mathcal{N}_X \quad (11b)$$

$$-M \cdot (1 - \xi_i) \leq F_i - 1 - \sum_{k=0}^{\lfloor \log_2 \lceil \hat{D}_{\text{dc},i} \rceil \rfloor} 2^k \cdot z_{k,i}^{(\lambda \cdot \Delta f)} \leq M \cdot (1 - \xi_i), \quad \forall i \in \mathcal{N}_X \quad (11c)$$

### 3) Timestamps for marking the phases

As shown in Fig. 2(a), three timestamps can be identified as (12a)–(12c), dividing the restored load into distinct phases: interruption, CLPU peak, CLPU decay, and post-CLPU steady load. Consequently, sets of binary variables are introduced as indicators of which phase is active during any given time span of the restoration process, denoted by

$$u_{\sim,i,t} = \begin{cases} 0, & \text{if } t \leq t_{\sim,i} \\ 1, & \text{if } t \geq t_{\sim,i} + 1 \end{cases}$$

Therefore, the above can be enforced by (12a)–(12f).

$$t_{1,i} = \sum_{t \in \mathcal{T}} (1 - \delta_{i,t}), \quad \forall i \in \mathcal{N}_X \quad (12a)$$

$$t_{2,i} = t_{1,i} + D_{\text{pk},i}, \quad \forall i \in \mathcal{N}_X \quad (12b)$$

$$t_{3,i} = t_{2,i} + D_{\text{dc},i}, \quad \forall i \in \mathcal{N}_X \quad (12c)$$

$$-M \cdot (1 - u_{1,i,t}) + 1 \leq t - t_{1,i} \leq M \cdot u_{1,i,t}, \quad \forall i \in \mathcal{N}_X, t \in \mathcal{T} \quad (12d)$$

$$-M \cdot (1 - u_{2,i,t}) + 1 \leq t - t_{2,i} \leq M \cdot u_{2,i,t}, \quad \forall i \in \mathcal{N}_X, t \in \mathcal{T} \quad (12e)$$

$$-M \cdot (1 - u_{3,i,t}) + 1 \leq t - t_{3,i} \leq M \cdot u_{3,i,t}, \quad \forall i \in \mathcal{N}_X, t \in \mathcal{T} \quad (12f)$$

### 4) Load representation

Based on the aforementioned efforts, we can derive the load including CLPU as

$$P_{L,i,t} = p_{L,i,t} \cdot \begin{Bmatrix} (u_{1,i,t} - u_{2,i,t}) \cdot F_i + (u_{2,i,t} - u_{3,i,t}) \cdot \\ [F_i - (t - t_{2,i} - 0.5) \cdot \Delta f_i] + u_{3,i,t} \end{Bmatrix}, \quad \forall i \in \mathcal{N}_X, t \in \mathcal{T} \quad (13)$$

There are bilinear terms and higher-dimensional products in (13). Nevertheless, most of the involved variables are binary or integer (notably, $u_{1,i,t} - u_{2,i,t}$ and $u_{2,i,t} - u_{3,i,t}$ are binary), facilitating the exact linearization. To achieve this, we introduce auxiliary variables $W_{1,i,t}$, $W_{2,i,t}$, and $W_{3,i,t}$ to substitute for $(u_{1,i,t} - u_{2,i,t}) \cdot F_i$, $(u_{2,i,t} - u_{3,i,t}) \cdot F_i$, and $(u_{2,i,t} - u_{3,i,t}) \cdot \Delta f_i$, respectively, with the following constraints added.

$$-\hat{F}_{i^*} \cdot (1 - u_{1,i,t} + u_{2,i,t}) \leq W_{1,i,t} - F_i \leq 0, \quad \forall i \in \mathcal{N}_X, t \in \mathcal{T} \quad (14a)$$

$$0 \leq W_{1,i,t} \leq \hat{F}_{i^*} \cdot (u_{1,i,t} - u_{2,i,t}), \quad \forall i \in \mathcal{N}_X, t \in \mathcal{T} \quad (14b)$$

$$-\hat{F}_{i^*} \cdot (1 - u_{2,i,t} + u_{3,i,t}) \leq W_{2,i,t} - F_i \leq 0, \quad \forall i \in \mathcal{N}_X, t \in \mathcal{T} \quad (14c)$$

$$0 \leq W_{2,i,t} \leq \hat{F}_{i^*} \cdot (u_{2,i,t} - u_{3,i,t}), \quad \forall i \in \mathcal{N}_X, t \in \mathcal{T} \quad (14d)$$

$$-M \cdot (1 - u_{2,i,t} + u_{3,i,t}) \leq W_{3,i,t} - \Delta f_i \leq 0, \quad \forall i \in \mathcal{N}_X, t \in \mathcal{T} \quad (14e)$$

$$0 \leq W_{3,i,t} \leq M \cdot (u_{2,i,t} - u_{3,i,t}), \quad \forall i \in \mathcal{N}_X, t \in \mathcal{T} \quad (14f)$$

After this, the only remaining bilinear term $t_{2,i} \cdot W_{3,i,t}$ can be further linearized using binary decomposition, as $t_{2,i}$ is an integer. With (15), $t_{2,i} \cdot W_{3,i,t}$ can be replaced with $W_{4,i,t}$.

$$t_{2,i} = \sum_{k=0}^{\lfloor \log_2 \bar{t}_2 \rfloor} 2^k \cdot \lambda_{k,i}^{(t)}, \quad \forall i \in \mathcal{N}_X \quad (15a)$$

$$-M \cdot (1 - \lambda_{k,i}^{(t)}) \leq z_{k,i,t}^{(\lambda \cdot W)} - W_{3,i,t} \leq 0, \quad \forall k \in \mathcal{K}_t, i \in \mathcal{N}_X, t \in \mathcal{T} \quad (15b)$$

$$0 \leq z_{k,i,t}^{(\lambda \cdot W)} \leq M \cdot \lambda_{k,i}^{(t)}, \quad \forall k \in \mathcal{K}_\alpha, i \in \mathcal{N}_X, t \in \mathcal{T} \quad (15c)$$

$$W_{4,i,t} = \sum_{k=0}^{\lfloor \log_2 \bar{t}_2 \rfloor} 2^k \cdot z_{k,i,t}^{(\lambda \cdot W)} \;,\; \forall i \in \mathcal{N}_X, t \in \mathcal{T} \quad (15d)$$

Similarly to (5), McCormick inequalities can be added:

$$W_{4,i,t} \geq \overline{W}_{3,i,t} \cdot t_{2,i} + \overline{t}_{2,i} \cdot W_{3,i,t} - \overline{t}_{2,i} \cdot \overline{W}_{3,i,t} \;,\; \forall i \in \mathcal{N}_X, t \in \mathcal{T} \quad (16)$$

where $\bar{t}_2$ and $\overline{W}_{3,i,t}$ are the estimated upper bounds for all the possible values of $t_{2,i}$ and $W_{3,i,t}$, respectively.

Finally, (13) can be rewritten as

$$P_{L,i,t} = p_{L,i,t} \cdot \left[ W_{1,i,t} + W_{2,i,t} - (t-0.5) \cdot W_{3,i,t} + W_{4,i,t} + u_{3,i,t} \right] \quad (17)$$

which provides a generalized linear formulation for the interrupted loads throughout the entire restoration horizon.

## IV. MODELING OF RESTORATION PROBLEM

### A. Routing and scheduling of MESS

As an effective and clean solution for providing emergency supply, MESS is integrated into our restoration process by optimizing its routing and scheduling. The routing model from the authors' previous work [39] is adopted, while the scheduling or operational constraints are expressed by

$$ch_{j,i,t} + dch_{j,i,t} \leq x_{j,i,t} \;,\; \forall j \in \mathcal{M}, i \in \mathcal{N}_M, t \in \mathcal{T} \quad (18a)$$

$$0 \leq P_{c,j,i,t} \leq ch_{j,i,t} \cdot P_{c,\max,j} \;,\; \forall j \in \mathcal{M}, i \in \mathcal{N}_M, t \in \mathcal{T} \quad (18b)$$

$$0 \leq P_{d,j,i,t} \leq dch_{j,i,t} \cdot P_{d,\max,j} \;,\; \forall j \in \mathcal{M}, i \in \mathcal{N}_M, t \in \mathcal{T} \quad (18c)$$

$$-x_{j,i,t} \cdot Q_{\max,j} \leq Q_{j,i,t} \leq x_{j,i,t} \cdot Q_{\max,j}, \forall j \in \mathcal{M}, i \in \mathcal{N}_M, t \in \mathcal{T} \quad (18d)$$

$$\left[ \sum_{i \in \mathcal{N}_M} \left( P_{d,j,i,t} - P_{c,j,i,t} \right) \right]^2 + \left( \sum_{i \in \mathcal{N}_M} Q_{j,i,t} \right)^2 \leq S_j^2 \quad (18e)$$
$$, \forall j \in \mathcal{M}, t \in \mathcal{T}$$

$$soc_{j,t} = soc_{j,t-1} + \left( e_{c,j} \cdot \sum_{i \in \mathcal{N}_M} P_{c,j,i,t} - \frac{\sum_{i \in \mathcal{N}_M} P_{d,j,i,t}}{e_{d,j}} \right) \cdot \frac{\Delta t}{E_j} \quad (18f)$$
$$, \forall j \in \mathcal{M}, t \in \mathcal{T}$$

$$soc_{j,\min} \leq soc_{j,t} \leq soc_{j,\max} \;,\; \forall j \in \mathcal{M}, t \in \mathcal{T} \quad (18g)$$

### B. Network reconfiguration

The flow-based model from [40] is used to formulate the network reconfiguration of PDS, facilitating the formation of microgrids to supply interrupted loads, as follows:

$$\sum_{(i',i_r)\in \mathcal{L}} f_{i'i_r,t}^{i_x} - \sum_{(i_r,i')\in \mathcal{L}} f_{i_r i',t}^{i_x} = -1 \;,\; \forall t \in \mathcal{T}, i_x \in \mathcal{N} \setminus i_r \quad (19a)$$

$$\sum_{(i',i_x)\in \mathcal{L}} f_{i'i_x,t}^{i_x} - \sum_{(i_x,i')\in \mathcal{L}} f_{i_x i',t}^{i_x} = 1 \;,\; \forall t \in \mathcal{T}, i_x \in \mathcal{N} \setminus i_r \quad (19b)$$

$$\sum_{(i',i)\in \mathcal{L}} f_{i'i,t}^{i_x} - \sum_{(i,i')\in \mathcal{L}} f_{ii',t}^{i_x} = 0$$
$$, \forall t \in \mathcal{T}, i_x \in \mathcal{N}\setminus i_r, i \in \mathcal{N}\setminus\{i_x,i_r\} \quad (19c)$$

$$0 \leq f_{i'i,t}^{i_x} \leq \lambda_{i'i,t} \;,\; 0 \leq f_{ii',t}^{i_x} \leq \lambda_{ii',t}$$
$$, \forall t \in \mathcal{T}, i_x \in \mathcal{N}\setminus i_r, (i',i) \in \mathcal{L} \quad (19d)$$

$$\sum_{(i',i)\in \mathcal{L}} \left( \lambda_{i'i,t} + \lambda_{ii',t} \right) = |\mathcal{N}| - 1 \;,\; \forall t \in \mathcal{T} \quad (19e)$$

$$\lambda_{i'i,t} + \lambda_{ii',t} = \mu_{i'i,t} \;,\; \forall t \in \mathcal{T}, (i',i) \in \mathcal{L} \quad (19f)$$

$$\upsilon_{i'i,t} \leq \mu_{i'i,t} \;,\; \forall t \in \mathcal{T}, (i',i) \in \mathcal{L} \quad (19g)$$

$$\upsilon_{i'i,t} = 0 \;,\; \forall (i',i) \in \mathcal{L}_{\text{out},t} \quad (19h)$$

where $i_r$ represents the substation node.

### C. PDS operation

The widely-used linearized DistFlow model is adopted as the operational constraints of the PDS [4], [6], incorporating backup distributed generators.

$$P_{\text{IN},i,t} = [\![ i \in \mathcal{N}_M ]\!] \cdot \sum_{j\in M} \left( P_{d,j,i,t} - P_{c,j,i,t} \right) + P_{\text{DG},i,t}$$
$$, \forall t \in \mathcal{T}, i \in \mathcal{N} \quad (20a)$$

$$Q_{\text{IN},i,t} = [\![ i \in \mathcal{N}_M ]\!] \cdot \sum_{j\in M} Q_{j,i,t} + Q_{\text{DG},i,t} \;,\; \forall t \in \mathcal{T}, i \in \mathcal{N} \quad (20b)$$

$$\sum_{(i',i)\in \mathcal{L}} P_{i'i,t} + P_{\text{IN},i,t} - P_{L,i,t} = \sum_{(i,i')\in \mathcal{L}} P_{ii',t} \;,\; \forall t \in \mathcal{T}, i \in \mathcal{N} \quad (20c)$$

$$\sum_{(i',i)\in \mathcal{L}} Q_{i'i,t} + Q_{\text{IN},i,t} - Q_{L,i,t} = \sum_{(i,i')\in \mathcal{L}} Q_{ii',t} \;,\; \forall t \in \mathcal{T}, i \in \mathcal{N} \quad (20d)$$

$$Q_{L,i,t} = \eta_i \cdot P_{L,i,t} \;,\; \forall t \in \mathcal{T}, i \in \mathcal{N}_X \quad (20e)$$

$$\begin{cases} V_{i,t}^2 \geq V_{i',t}^2 - 2\left( P_{i'i,t} r_{i'i} + Q_{i'i,t} x_{i'i} \right) - M\left( 1 - \upsilon_{i'i,t} \right) \\ V_{i,t}^2 \leq V_{i',t}^2 - 2\left( P_{i'i,t} r_{i'i} + Q_{i'i,t} x_{i'i} \right) + M\left( 1 - \upsilon_{i'i,t} \right) \end{cases} \quad (20f)$$
$$, \forall t \in \mathcal{T}, (i',i) \in \mathcal{L}$$

$$V_{i,\min}^2 \leq V_{i,t}^2 \leq V_{i,\max}^2 \;,\; \forall t \in \mathcal{T}, i \in \mathcal{N} \quad (20g)$$

$$P_{i'i,t}^2 + Q_{i'i,t}^2 \leq \upsilon_{i'i,t} \cdot S_{i'i}^2 \;,\; \forall t \in \mathcal{T}, (i',i) \in \mathcal{L} \quad (20h)$$

$$\delta_{i,t} \geq \delta_{i,t-1} \;,\; \forall t \in \mathcal{T}, i \in \mathcal{N}_X \quad (20i)$$

$$P_{\text{DG},i,t}^2 + Q_{\text{DG},i,t}^2 \leq S_{\text{DG},i}^2 \;,\; \forall t \in \mathcal{T}, i \in \mathcal{N} \quad (20j)$$

### D. Objective function for restoration

The proposed strategy aims to restore the interrupted loads with the objective of minimizing the total CIC. In addition, costs associated with deploying MESS, operating distributed generators, and switching branches are included to avoid futile MESS movements, generator operations, and switching actions. The objective function is thus formulated as

$$\begin{aligned} \min \;\; & \sum_{t\in \mathcal{T}} \sum_{i\in \mathcal{N}_X} p_{L,i,t} \cdot A_{i,t}^{\text{fun}} \cdot \Delta t \\ & + \sigma_1 \cdot \sum_{t\in \mathcal{T}} \sum_{m\in \mathcal{M}} \sum_{i\in \mathcal{N}_M} v_{j,i,t} \\ & + \sigma_2 \cdot \sum_{t\in \mathcal{T}} \sum_{i\in \mathcal{N}} P_{\text{DG},i,t} \\ & + \sigma_3 \cdot \sum_{t\in \mathcal{T}} \sum_{(i',i)\in \mathcal{L}} \left| \upsilon_{i'i,t} - \upsilon_{i'i,t-1} \right| \end{aligned} \quad (21)$$

Given that $|\upsilon_{i'i,t} - \upsilon_{i'i,t-1}|$ can be equally written as "max $\{\upsilon_{i'i,t} - \upsilon_{i'i,t-1}, \upsilon_{i'i,t-1} - \upsilon_{i'i,t}\}$", we introduce an auxiliary variable $\Delta \upsilon_{i'i,t}$ to replace $|\upsilon_{i'i,t} - \upsilon_{i'i,t-1}|$ with the following constraints imposed, thereby converting (21) into a linear form.

$$\Delta \upsilon_{i'i,t} \geq \upsilon_{i'i,t} - \upsilon_{i'i,t-1} \;,\; \Delta \upsilon_{i'i,t} \geq \upsilon_{i'i,t-1} - \upsilon_{i'i,t} \quad (22)$$

## V. CASE STUDIES

Case studies are conducted to validate the effectiveness of the proposed strategy. While the cone constraints (18e), (20h), and (20j) can be linearized using the method from [41], the mixed-integer linear programming (MILP) model for PDS restoration optimization, formulated in Section II to IV, is programmed in MATLAB using the YALMIP toolbox [42] and solved by Gurobi v11.0.0.

### A. Test System

IEEE 33-node system is used for test and illustrated in Fig. 3, with detailed parameters available in [43]. We set up initial damages on 7 branches to construct the testing scenario. Since this paper focuses on modeling CIC and CID rather than detailing a resource-intensive PDS restoration, the repairs for these damaged branches are assumed predefined and occurs in two stages, as shown in Table 1. The scheduling horizon is set to 9 hours, with each time span defined as $\Delta t$=30 minutes.

The loads are arbitrarily divided into residential and C&I types, with variations in their CIC relative to CID modeled according to the parameters in Fig. 1. The initial CID $\alpha_{i,0}$ is uniformly set to 1 – corresponding to an initial interruption of 30 minutes, except for the nodes continuously connected to the substation node 1. The relationships between CLPU characteristics and CID, as shown in Fig. 4, are assumed for CLPU modeling. In practice, these relationships can be derived from simulated or historical CLPU data under various CID scenarios. However, as this lies outside the scope of this paper, we make a direct assumption here. Details on relevant simulation methodologies can be found in literature such as [26]-[28], [44].

Three 500kW/1000kW·h MESS units are dispatched during the restoration process. Three MESS connection points are assumed at nodes 3, 8, and 29 within the PDS. The travel time between nodes 3 and 29 is set to 2 time spans, while for other node pairs, it is set to 1 time span. The MESS units are initially located at node 3 with an initial SOC of 0.9. In addition, two 800kW/1000kVA generators are available at nodes 11 and 24.

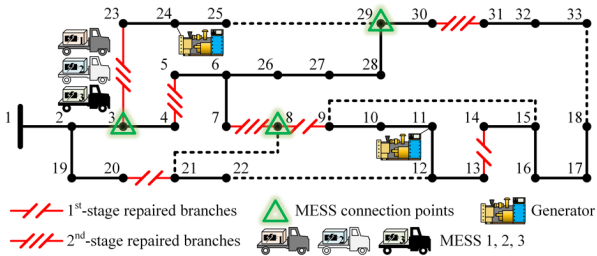

**Fig. 3.** IEEE 33-node system for test.

TABLE I
REPAIR OF DAMAGED BRANCHES

| Stages | Set of damaged branches: $\mathcal{L}_{\text{out},t}$ |
|---|---|
| 1st to 7th time spans | (20, 21), (8, 9), (13, 14), (3, 23), (4, 5), (7, 8), (30, 31) |
| 8th to 11th time spans | (3, 23), (4, 5), (7, 8), (30, 31) |

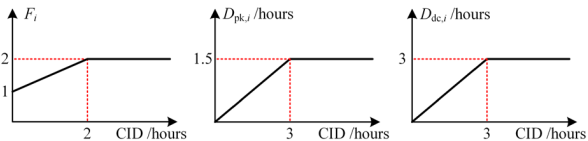

**Fig. 4.** Assumed relationships between CLPU characteristics and CID.

### B. Results Analysis

The main results are presented as follows. First, we examine the overall restoration process, as illustrated in Fig. 5. Under the fault scenario described in Table I, the PDS is initially reconfigured into two load islands, which are subsequently reconnected to the main grid as the damaged branches are gradually repaired after the 7th and 11th time spans. During this process, the network reconfiguration requires only two necessary steps: closing branch 20-21 after the 7th time span and closing branch 3-23 after the 11th time span. Other damaged branches, even after repair, do no need further operation to maintain the radial structure of the PDS and minimize the switching actions, as dedicated by the last term of the objective function (21).

The three MESSs are deployed to the two islands at the start of the restoration process, as shown in Figs. 5-6. Notably, MESS 1 departs from Island 2 before it reconnects to the main grid. This is because the CLPU process of the restored loads within Island 2 has already subsided, as shown in Fig. 8, and the load has dropped to a level that the generator at node 11 can sustain by itself, giving MESS 1 the opportunity to leave and supply another island. Then, as load 11 is restored during the 7th time span, the generator alone would be insufficient to sustain Island 2; consequently, MESS 3 arrives to provide supplementary power. Once Island 2 reconnects to the grid, MESS 3 and MESS 2, which has just arrived due to proximity to node 8, charge and then return to Island 1 to supply power.

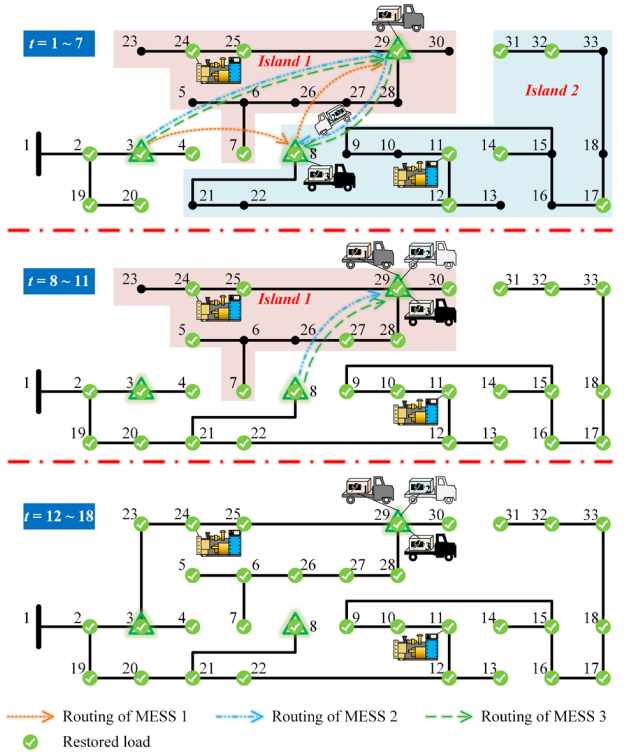

**Fig. 5.** Results of the PDS restoration process (each subfigure shows the state during the final time span of each period).

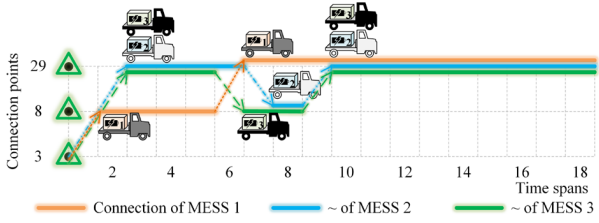

**Fig. 6.** Results of routing of MESSs.

After reviewing the overall restoration process, we now turn our attention to the results for CIC and CLPU. The detailed results in Figs. 7-9 clearly demonstrate the effectiveness of the proposed model for capturing the decision-dependency of CLPU. Specifically, Figs. 7(a) and 7(b) present the results of the increasing rate of CIC – $A_{i,t}^{\text{fun}}$, with the current case referred to as the *base case*. For clearer demonstration, we also provide the results for a case with an increased initial CID, shown in Figs. 7(c) and 7(d). These findings indicate that the proposed model effectively reproduces the relationship between CIC and CID depicted in Fig. 1, offering a precise and objective approach for quantifying CIC and defining objectives in restoration optimization modeling.

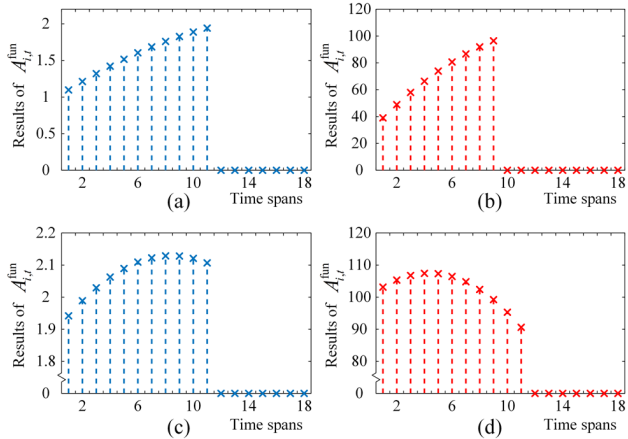

**Fig. 7.** Results of increasing rates of CIC for (a) node 23 and (b) node 30 under the *base case*; and (a) node 23 and (b) node 30 under the case of an initial interruption of 5.5 h.

Fig. 8 presents the results of load states, with particular emphasis on the distribution of different phases in the post-restoration CLPU process for all loads, revealing the anticipated variations in CLPU timing characteristics across different CID conditions. To further highlight our findings, we showcase several representative load curves in Fig. 9 to more clearly illustrate the variations in CLPU, where the loads have been normalized by their original values without CLPU. The results show that as CID gradually increases, the CLPU peak magnitude, peak duration, and decay duration all increase correspondingly; once CID surpasses the respective saturation points, these characteristics stabilize at their maximum values. Collectively, these findings from the results demonstrate that the proposed model effectively captures and reproduces the relationships between the key CLPU characteristics and CID, as pre-defined in Fig. 4. This enables a dynamic representation of CLPU relative to CID, enhancing the precision in modeling load restoration behaviors. In turn, it enables the efficient deployment of emergency resources and facilitates more effective restoration strategies. Evidently, without such dynamic representation, CLPU for certain periods may be overestimated or underestimated, resulting in excessive deployment of "larger-than-necessary" resources or rendering the "optimal" decisions practically infeasible.

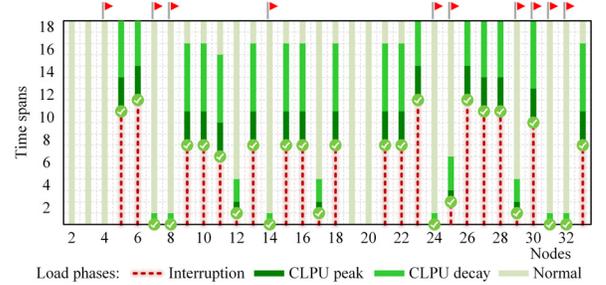

**Fig. 8.** Results of load states (where the flag symbol denotes the C&I customers).

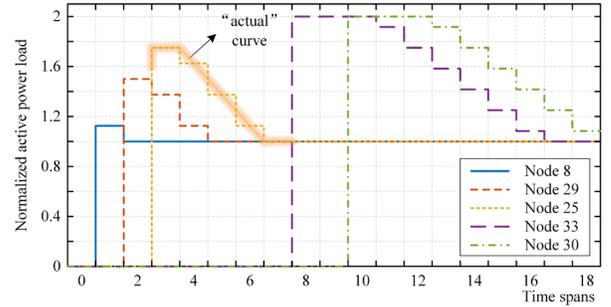

**Fig. 9.** Typical restored load curves incorporating decision-dependent CLPU.

### C. "Dynamic Rating" by MESS

To further reveal the potential of MESS in enhancing restoration performance, particularly under the influence of CLPU, we present the power demand and supply results for the two Islands shown in Fig. 5, as illustrated in Fig. 10 (a). Since the two Islands reconnect to the main grid after the 7[th] time span and 11[th] time span, respectively, only the profiles before these periods are plotted.

For comparison, we also simulate two additional cases: one where the mobility of MESSs is removed (by fixing their locations within the two Islands), and another where all MESSs are entirely removed. The results are shown in Fig. 10 (b) and Fig. 10 (c), respectively.

From Fig. 10 (a), it can be observed that the generators serve as the primary power supply, nearly always operating at their maximum capacity to support the restored loads. However, the CLPU following load restoration often drives power demand beyond the generator's capability. In such situations, MESS provides short-term support to address this excessive power demand. Once the CLPU subsides and the demand falls back within the generator's capacity, the MESS can leave, as clearly demonstrated by the behavior of MESS 1 in Island 2.

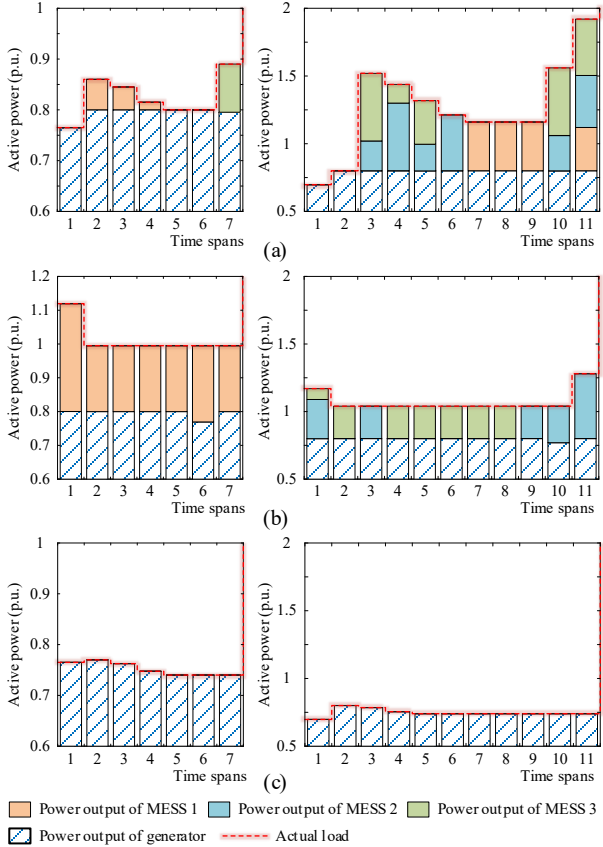

**Fig. 10.** Results of power demand and scheduled generation for Island 2 (left) and Island 1 (right) under (a) the *base case*, (b) the case without mobility, and (c) the case without MESS. The load is represented in p.u. with a base power of 1 MW.

This observation underscores the pivotal role of MESS as a provider of "dynamic rating" service during the PDS restoration process – a term originally coined in the context of transmission lines to describe the temporary boost in transmission capacity to deliver more power when needed. Similarly, during restoration, MESS temporarily augments the total supply capacity of isolated islands, enabling them to withstand temporary load surges, such as those induced by CLPU. Without this temporary support, the amount of restored load would have to be reduced to limit the load surges, as shown in Fig. 10 (c). Furthermore, the mobility of MESS facilitates the flexibility of this dynamic rating service. As observed in the previous results, once the CLPU-induced load surges subside, the additional capacity provided by MESS can be released, allowing the MESS to relocate to other islands and continue supporting their restoration efforts. For a more intuitive comparison, the total CIC for the base case and the cases of Fig. 10 (b) and (c) are $ 91526, $ 134760, and $ 271340, respectively. This indicates that the utilization of MESS reduces the total CIC by 32 % and 66 %, compared to the cases without mobility and without MESS, respectively.

## VI. CONCLUSION

Incorporating the decision-dependencies of CIC and CLPU into PDS restoration decision-making is critical for developing effective and objective restoration strategies, which remains challenging due to the unclear physical mechanisms governing these dependencies. To address this, inspired by the principles of data-driven optimization, this paper proposes tractable modeling approaches to formulate the decision-dependent CIC and CLPU, thereby capturing and representing the underlying varying patterns of CIC and CLPU with CID from data. Specifically, all related constraints are derived in linear forms, based on which a comprehensive PDS restoration optimization problem is constructed. The results of the case studies demonstrate the effectiveness of the proposed modeling approaches. In addition, the impressive role of MESS in providing a "dynamic rating" service during the CLPU-related restoration process is highlighted, showcasing its potential to enhance the restoration performance.

This study serves as a stepping stone toward addressing the decision-dependencies of CIC and CLPU in PDS restoration. Future research could focus on developing fully analytical models, grounded in a deeper investigation of the underlying mechanisms behind these decision-dependencies, to improve interpretability, or refining the modeling techniques to construct more precise representations within a data-driven optimization framework.